\documentstyle[11pt,newpasp,twoside,epsf]{article}
\def\edcomment#1{\iffalse\marginpar{\raggedright\sl#1\/}\else\relax\fi}
\reversemarginpar

\begin{document}
\title{The Light Elements Be and B as Stellar Chronometers in the Early Galaxy}
 \author{Timothy C. Beers}
\affil{Michigan State University, Dept. of Physics \& Astronomy, E. Lansing, MI
 48824  USA}
\author{Takeru K. Suzuki}
\affil{University of Tokyo, Dept. of Astronomy, School of Science,
University of Tokyo; Theoretical
Astrophysics Division, National 
Astronomical Observatory, Mitaka, Tokyo, 181-8588 Japan}
\author{Yuzuru Yoshii}
\affil{University of Tokyo, Institute of Astronomy, School of Science, 
University of Tokyo, Mitaka, Tokyo, 181-8588 Japan; 
Research Center for the Early Universe, School 
of Science, University of Tokyo Japan}

\begin{abstract}
Recent detailed simulations of Galactic Chemical Evolution have shown that the
heavy elements, in particular [Fe/H], are expected to exhibit a weak, or
absent, correlation with stellar ages in the early Galaxy due to the lack of
efficient mixing of interstellar material enriched by individual Type II
supernovae.  A promising alternative ``chronometer'' of stellar ages is
suggested, based on the expectation that the light elements Be and B are formed
primarily as spallation products of Galactic Cosmic Rays.
\end{abstract}

\section{Introduction}

It has become clear, from a number of lines of recent evidence, that the early
evolution of the Galaxy is best thought of as a stochastic process.   Within
the first 0.5-1 Gyr following the start of the star formation process, chemical
enrichment does not operate within a well-mixed uniform environment, as was
assumed in the simple one-zone models that were commonly used in past
treatments of this problem.  Rather, the very first generations of stars are
expected to have their abundances of heavy elements set by local conditions,
which are likely to have been dominated by the yields from individual SNeII.

The seeds of this paradigm shift can be found in the observations,
interpretations, and speculations of McWilliam et al. (1995), Audouze \& Silk
(1995), and Ryan, Norris, \& Beers (1996).  Models which attempt to incorporate
these ideas into a predictive formalism have been put forward by Tsujimoto,
Shigeyama, \& Yoshii (1999; hereafter TSY), and Argast et al.  (2000).
Although they differ in the details of their implementation, and in a number of
their assumptions, both of these models rely on the idea of enhanced star
formation in the high-density shells of SN remnants, and the interaction of
these shells of enriched material with a local ISM.  The predictions which
result are similar as well: (1) Both models are capable of reproducing the
observed distributions of abundance (e.g., [Fe/H]) for stars in the tail of the
halo metallicity distribution function (Laird et al. 1988; Ryan \& Norris 1991;
Beers 1999), and (2) Both models predict that the abundances of heavy elements,
such as Fe, are not expected to show strong correlations with the ages of the
first stars, at least up until an enrichment level on the order of [Fe/H]$ \sim
-2.0$ is reached, i.e., at the time when mixing on a Galactic scale is possible
(roughly 1 Gyr following the initiation of star formation).

Suzuki, Yoshii, \& Kajino (1999; hereafter SYK, see also Suzuki, Yoshii, \&
Kajino, this volume) have extended the SN-induced chemical evolution model of
TSY to include predictions of the evolution of the light element species
$^9$Be, $^{10}$B, and $^{11}$B, based on secondary processes involving
spallative reactions with Galactic Cosmic Rays (hereafter GCRs).  Recently,
Suzuki, Yoshii, \& Beers (2000) have considered the extension of this model to
the prediction of $^6$Li and $^7$ Li, and demonstrate that they naturally
reproduce the recently detected slope in the abundance of Li in extremely
metal-poor stars noted by Ryan, Norris, \& Beers (1999; see also Ryan this
volume).  It is particularly encouraging that the same stochastic
star-formation models which reproduce the observed trends of some (but not all)
heavy elements, such as Eu, Fe, etc., also obtain predictions of the light
element abundance distributions that match the available observations quite
well, with a minimum of parameter tweaking.

In this contribution we summarize one of the more interesting predictions of
the TSY/SYK class of models, that the abundances of the light elements Be and B
(hereafter, BeB) might be useful as stellar chronometers in the early Galaxy (a
time when the heavy element ``age-metallicity'' relationships are not operating
due to the lack of global mixing).  It appears possible that, with refinement
of the modeling, and adequate testing, observations of BeB for metal-poor
stars may provide a chronometer with ``time resolution'' on scales of tens
of Myrs.
		
\section{The Essence of the Model}

In this section we would like to briefly explain our model of SN-induced
star formation and chemical evolution.  After formation of the very FIRST
generation of (Pop. III) stars, with atmospheres containing gas of primordial
abundance, the most massive of these stars exhaust their core H, and explode as
SNeII.  Following the explosion a shock is formed, because the velocity of the
ejected material exceeds the local sound speed.  Behind the shock the
swept-up ambient material in the ISM accumulates to form a high-density shell.
This shell cools in the later stages of the lifetime of a given SN remnant
(SNR) and is a suitable site for the star formation process to occur.  The
SNR shells are expected to be distributed randomly throughout the early and
rapidly evolving halo, and the shells do not easily merge with one another
because of the large available volume.  As a result, each SNR keeps its
identity and the stars which form there reflect the abundances of material
generated by their ``parent'' SN.  TSY present this model, and describe the
input assumptions, in more quantitative detail.  Figure 1 provides a cartoon
illustration of the processes which we discuss herein.

\begin{figure}
%\epsfxsize=11cm
%\epsfysize=13cm
%\epsfbox{beersf1.eps}
\plotfiddle{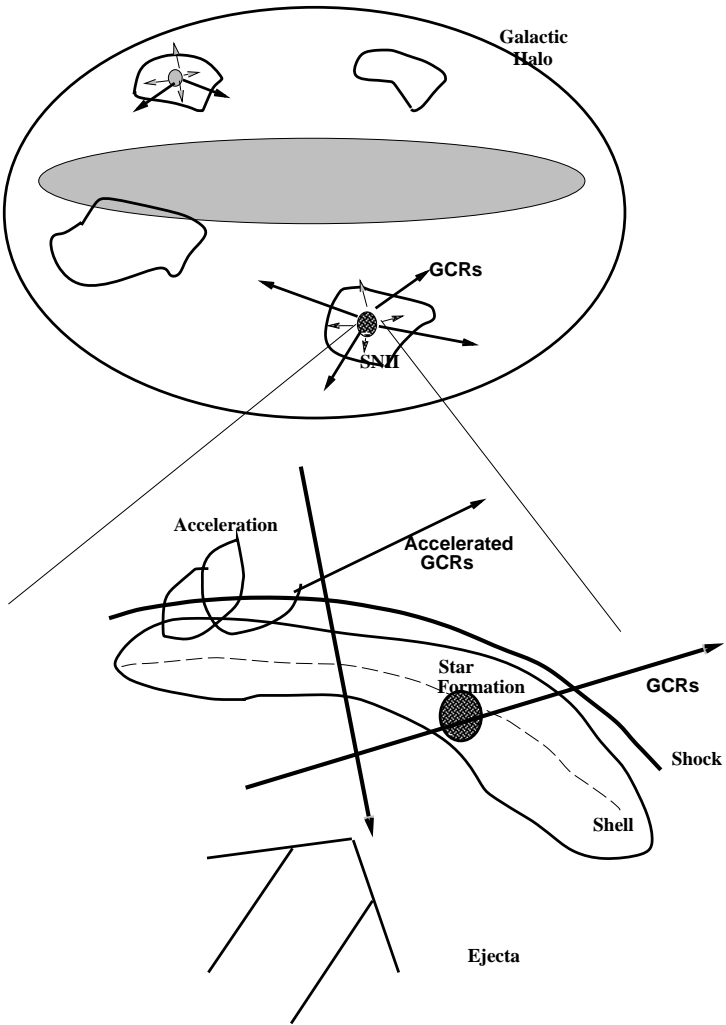}{11cm}{0.}{65.}{65.}{-215.}{-33.}
\caption{A simplified view of the early stages of chemical evolution in 
the Galactic halo. In the lower cutout we show star formation 
being triggered in SNR shells. See text for more detail.} 
\end{figure}

One of the most important results of the TSY model is that stellar metallicity,
especially [Fe/H], cannot be employed as an age indicator at these early
epochs.  Thus, to consider the expected elemental abundances of the metal-poor
stars which form at a given time, a {\it distribution} of stellar abundances
must be constructed, rather than adopting a global average abundance under the
assumption that the gas of the ISM is well mixed.  SYK constructed such a
model, coupled with the model of SN-induced chemical evolution, which considers
the evolution of the light elements.

SYK proposed that GCRs arise from the mixture of elements of individual SN
ejecta and their swept-up ISM, with the acceleration being due to the shock
formed in the SNR.  GCRs originating from SNeII propagate faster than the
material trapped in the clouds of gas making up the early halo.  As a result,
GCRs are expected to achieve uniformity throughout the halo faster than
the general ISM, with its patchy structure.  It follows that the abundances of
BeB, which are mainly produced by spallation processes of CNO elements
involving GCRs, are expected to exhibit a much tighter correlation with time
than those of heavy elements, synthesized through stellar evolution and SN
explosions.

We note that alternative models for the origin of spallative nucleosynthesis
products have been developed which rely on the existence of {\it spatially
correlated} SNeII in superbubbles of the early ISM (see Parizot \& Drury
1999, and this volume).  The superbubble model predicts a locally homogeneous
production of both heavy and light elements, and the variety of stellar
abundances which are observed are explained by the differing diffusion
processes of metal-rich ([Fe/H] $\sim -1$) shells swept-up by the bubble and
mixed with a metal-poor ([Fe/H] $\sim -4$) ISM.  Tests of the ``isolated'' SN
models vs. the superbubble models are expected to be conducted in the near
future. 
  
\section{Abundance Predictions of the Model}

Figure 2 shows the predicted behavior of the abundance of [Fe/H], log(Be/H),
and log(B/H), as a function of time, over the first 0.6 Gyrs of the evolution
of the early Galaxy, according to the model of SYK.  At any given time (note
that ``zero time'' is set by the onset of star formation, not the beginning of
the Universe) the range of observed BeB is substantially less than that of Fe,
owing to the global nature of light element production.  For example, at time
0.2 Gyrs, the expected stellar [Fe/H] extends over a range of 50, while that of
log (BeB/H) is on the order of 3--7.

During early epochs Fe is produced {\it only} by SNeII, and most of the Fe
observed in stars formed in SNR shells originates from that contributed by the
parent SN, because of uniformly low Fe abundance in the ISM at that time.
Thus, the expected [Fe/H] of stars born at that time will exhibit a rather
large range, reflecting differences in Fe yields associated with the different
masses of the progenitor stars.  On the other hand, according to the SYK model,
most of the BeB is produced by spallation reactions of CNO nuclei involving
globally transported GCRs.  The observed abundances of BeB in metal-poor stars
which formed at this time should reflect the global nature of their production,
and the correlation between time and BeB abundance is expected to be much
better than that found for heavier species.

\begin{figure}
%\epsfxsize=15cm
%\epsfysize=6cm
%\epsfbox{beersf2.eps}
\plotfiddle{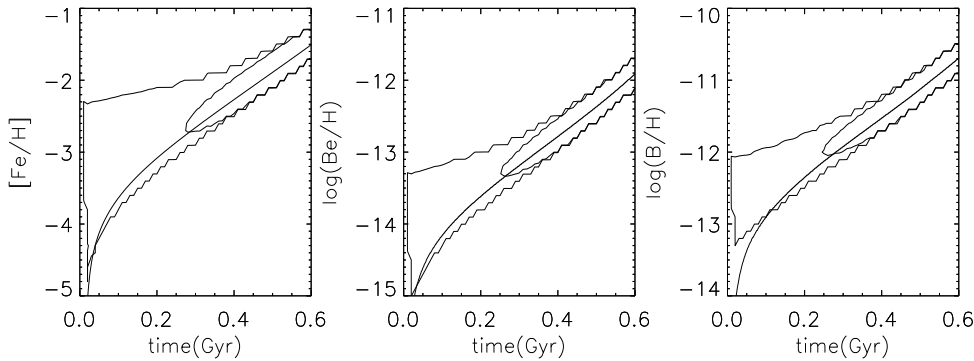}{5cm}{0.}{82.}{85}{-234.}{-55.}
\caption{Predicted distribution of abundance for three elements, relative to H,
for long-lived stars born at the indicated time {\it following} the initiation
of star formation.  The distributions have been convolved with Gaussians  with
$\sigma = 0.15$ dex to take into account expected observational errors.  The
two contours, from the inside to the outside, correspond to probability density
$10^{-3}$ and $10^{-5}$ within the unit area $\Delta t = 10({\rm Myr})\times
\Delta \log {\rm (element/H)} = 0.1$.  The solid lines show the predicted ISM
gas abundances of each element.}
\end{figure}

In Table 1, we use the predictions from SYK, and the stellar abundance data
from Boesgaard et al. (1999) for Be, to put forward ``bold'' estimates of
stellar ages (since the onset of star formation).  We note that these numbers
are meant to be indicative, not definitive, predictions, as further tests of
the model and its underlying assumptions still remain to be carried out.  We
have ordered the table according to estimated (Be) time since the onset of star
formation in the early Galaxy.

\begin{table}
\begin{center}
\caption{Predictions of Stellar ``Ages'' Based on Be Abundance}
\begin{tabular}{lccc}
\tableline
Star		&    [Fe/H] 	&  log(Be/H)	&   Be ``age'' (Gyr)    \cr
\tableline
BD$-$13:3442	&     $-$3.02	&   $-$13.49   &	0.22 ($-$0.07,+0.03) \cr 
BD+03:740	&     $-$2.89	&   $-$13.33   &	0.26 ($-$0.05,+0.03) \cr
HD 140283	&     $-$2.56	&   $-$13.08   &	0.32 ($-$0.05,+0.02) \cr 
BD+37:1458	&     $-$2.14	&   $-$13.07   &	0.32 ($-$0.05,+0.02) \cr
HD 84937	&     $-$2.20	&   $-$12.94   &	0.35 ($-$0.05,+0.03) \cr
BD+26:3578	&     $-$2.32	&   $-$12.79   &	0.39 ($-$0.05,+0.03) \cr
BD+02:3375	&     $-$2.39	&   $-$12.80   &	0.39 ($-$0.05,+0.03) \cr
BD$-$04:3208	&     $-$2.35	&   $-$12.69   &	0.41 ($-$0.05,+0.03) \cr
HD 19445	&     $-$2.10	&   $-$12.55   &	0.45 ($-$0.04,+0.03) \cr
HD 64090	&     $-$1.77	&   $-$12.49   &	0.46 ($-$0.04,+0.03) \cr
BD+20:3603	&     $-$2.22	&   $-$12.47   &	0.46 ($-$0.04,+0.03) \cr
BD+17:4708	&     $-$1.81	&   $-$12.40   &	0.48 ($-$0.04,+0.03) \cr
HD 219617	&     $-$1.58	&   $-$12.15   &	0.54 ($-$0.04,+0.02) \cr
HD 74000	&     $-$2.05	&   $-$12.10   &	0.55 ($-$0.04,+0.02) \cr
HD 103095	&     $-$1.37	&   $-$12.04   &	0.56 ($-$0.04,+0.02) \cr
HD 194598	&     $-$1.25	&   $-$11.88   &	0.59 ($-$0.03,+0.01) \cr
BD+23:3912	&     $-$1.53	&   $-$11.92   &	0.59 ($-$0.03,+0.01) \cr
HD 94028	&     $-$1.54	&   $-$11.55   &	$> 0.60$	     \cr  
\tableline
\tableline
\end{tabular}
\end{center}
\end{table}

It is interesting to consider the implications of this strong age-abundance
relationship for individual stars which have been noted in the literature as
having ``peculiar'' BeB (or $^7$Li for that matter) abundances, at least as
compared to otherwise similar stars of the same [Fe/H], T$_{\rm eff}$, and log
g.  The set of ``twins'' G64-12 and G64-37 have been noted as one example of
stars with very low metallicity, and apparently similar T$_{\rm eff}$ and log
g, which never-the-less, exhibit rather different abundances of $^7$Li.  Could
this difference be accounted for by a difference in AGE of these stars ?
Answering this question is of great importance, and hopefully will be resolved
in the near future.

\section{Can we Test This Model ?}

Yes, but it will take some hard work.  Obviously, if there exists an
independent method with which to verify the relative age determinations
predicted by this model, that would be ideal.  Fortunately, there have been
numerous refinements in models of stellar atmospheres, and their
interpretation, which may make this feasible (see Fuhrmann 2000).  In order to
apply the methods described by Fuhrmann, one requires high-resolution, high-S/N
spectroscopy of individual stars.  It is imperative that the present-generation
8m telescopes (VLT, SUBARU, GEMINI, HET) obtain this data, so that this, and
other related questions, may be addressed with the best possible information.

Another feasible test would be to compare the abundances of BeB with [Fe/H],
and other heavy elements, for a large sample of stars with [Fe/H] $< -2.0$.  If
the superbubble model is the correct interpretation, with an implied
locally homogeneous production of the light elements, then one might expect to
find correlations between the abundances of various heavy element species
(including those other than Fe and O) and BeB.  Simultaneous observations of
light and heavy elements for stars of extremely low abundance are planned with
all the major 8m telescopes, so it should not be too long before a sufficiently
large sample to carry out this test is obtained.

One can also seek, as we have, confirmatory evidence in the predicted behavior
of $^7$Li vs. [Fe/H] (Suzuki et al. 2000).

\section{Other Uses for This Model}

If the model we have considered here can be shown to be correct, there are
several new avenues of investigation which are immediately opened.
For example, if one were able to ``age rank'' stars on the basis of their BeB
abundances, one could refine alternative production mechanisms for the light
element Li which are not driven by GCR spallation, including the SN
$\nu-$process and/or production via a giant-branch Cameron-Fowler mechanism
(see Castilho et al. , this volume), in stellar  flares, etc..

Furthermore, since BeB nuclei are more difficult to burn than Li nuclei, one
could imagine a powerful test for the extent to which depletion of Li has
operated in metal-poor dwarfs, with important implications for the Li
constraint on Big Bang Nucleosynthesis (BBN).  Realistic modeling of BeB
evolution at early epochs may also help distinguish between predictions of
standard BBN, non-standard BBN, and the accretion hypothesis (see Yoshii,
Mathews, \& Kajino 1995).

An age ranking of metal-poor stars based on their BeB abundances, in
combination with measurements of their alpha, iron-peak, and neutron-capture
elements, would open the door for an unraveling of the mass spectrum of the
progenitors of first generation SNeIIs, and allow one to obtain direct
constraints on their elemental yields as a function of mass, a key component to
models of early nucleosynthesis.

\acknowledgements

TCB expresses gratitude to the IAU for support which enabled his attendance at
this meeting, and acknowledges partial support from the National Science
Foundation under grant AST 95-29454.   TCB also wishes to express his
congratulations to the LOC and SOC for a well-run, scientifically stimulating,
and marvelously located meeting.  YY acknowledges a Grant-in-Aid from the
Center of Excellence (COE), 10CE2002, awarded by the Ministry of Education,
Science, and Culture, Japan.


\begin{references}

\reference Argast D., Samlund, M., Gerhard, O.E., \& Thielemann, F.-K. 2000,
\aap, in press
\reference Audouze J., \& Silk, J. 1995, \apj , 451, L49 
\reference Beers, T.C. 1999, in Third Stromlo Symposium:  The Galactic Halo,
eds. B. Gibson, T. Axelrod, \& M. Putman, (ASP, San Francisco), 165, p. 206 
\reference Boesgaard, A.M., Deliyannis, C.P., King, J.R., Ryan, S.G., Vogt,
S.S., \& Beers, T.C. 1999, \aj , 117, 1549
\reference Fuhrmann, K. 2000, in The First Stars, Proceedings of the Second
MPA/ESO Workshop, eds. A. Weiss, T. Abel, \& V. Hill (Springer, Heidelberg), in
press
\reference Laird, J.B., Carney, B.W., Rupen, M.P., \& Latham, D.W. 1988, \aj, 
96, 1908
\reference McWilliam, A., Preston, W., Sneden, C., \& Searle, L. 1995, \aj ,
109, 2757
\reference Ryan, S.G., Norris, J.E., \& Beers, T.C. 1996, \apj , 471, 254
\reference Ryan, S.G., Norris, J.E., \& Beers, T.C. 1999, \apj, 523, 654 
\reference Ryan, S.G., \& Norris, J.E. 1991, \aj, 101, 1865
\reference Suzuki, T.K., Yoshii, Y., \& Kajino, T. 1999, \apj , 522, L125 
(SYK)
\reference Suzuki, T.K., Yoshii, Y., \& Beers, T.C. 2000, \apjl , submitted
\reference Tsujimoto, T., Shigeyama, T., \& Yoshii, Y. 1999, \apj , 519,
L63 (TSY) 
\reference Yoshii, Y., Mathews, G.J., \& Kajino, T. 1995, \apj, 447, 184
\end{references}
\end{document}